\documentclass[prl,reprint,superscriptaddress,longbibliography,twocolumn]{revtex4-1}

\usepackage{graphicx} 
\usepackage{amsmath,amssymb} 
\usepackage{bm} 
\usepackage{color} 
\usepackage{hyperref}

\hypersetup{
colorlinks=true,
urlcolor= blue,
citecolor=blue,
linkcolor= blue,
bookmarks=true,
bookmarksopen=false,
}

\renewcommand{\vec}[1]{\bm{#1}}

\begin{document}
\title{Unexpected Zero Bias Conductance Peak on the Topological Semimetal Sb(111) with a Broken Bilayer}

\author{Yau-Chuen Yam}
\affiliation{Department of Physics and Astronomy, University of British Columbia, Vancouver, British Columbia V6T 1Z4, Canada}
\affiliation{Department of Physics, Harvard University, Cambridge, Massachusetts 02138, USA}

\author{Shiang Fang}
\affiliation{Department of Physics, Harvard University, Cambridge, Massachusetts 02138, USA}

\author{Pengcheng Chen}
\affiliation{Department of Physics and Astronomy, University of British Columbia, Vancouver, British Columbia V6T 1Z4, Canada}
\affiliation{Department of Physics, Harvard University, Cambridge, Massachusetts 02138, USA}

\author{Yang He}
\affiliation{Department of Physics, Harvard University, Cambridge, Massachusetts 02138, USA}

\author{Anjan Soumyanarayanan}
\affiliation{Department of Physics, Harvard University, Cambridge, Massachusetts 02138, USA}

\author{Mohammad Hamidian}
\affiliation{Department of Physics, Harvard University, Cambridge, Massachusetts 02138, USA}

\author{Dillon Gardner}
\affiliation{Department of Physics, Massachusetts Institute of Technology, Cambridge, MA 02139, USA
}
\author{Young Lee}
\affiliation{Department of Physics, Massachusetts Institute of Technology, Cambridge, MA 02139, USA
}

\author{Marcel Franz}
\affiliation{Department of Physics and Astronomy, University of British Columbia, Vancouver, British Columbia V6T 1Z4, Canada}

\author{Bertrand I. Halperin}
\affiliation{Department of Physics, Harvard University, Cambridge, Massachusetts 02138, USA}

\author{Efthimios Kaxiras}
\affiliation{Department of Physics, Harvard University, Cambridge, Massachusetts 02138, USA}

\author{Jennifer E. Hoffman}
\affiliation{Department of Physics and Astronomy, University of British Columbia, Vancouver, British Columbia V6T 1Z4, Canada}
\affiliation{Department of Physics, Harvard University, Cambridge, Massachusetts 02138, USA}

\date{\today}

\begin{abstract}
The long-sought Majorana fermion is expected to manifest in a topological-superconductor heterostructure as a zero bias conductance peak (ZBCP). As one promising platform for such heterostructures, we investigate the cleaved surface of the topological semimetal Sb(111) using scanning tunneling microscopy and spectroscopy.
Remarkably, we find a robust ZBCP on some terraces of the cleaved surface, although no superconductor is present.
Using quasiparticle interference imaging, Landau level spectroscopy and density functional theory, we show that the ZBCP originates from a van Hove singularity pushed up to the Fermi level by a sub-surface stacking fault.
Amidst the sprint to stake claims on new Majorana fermion systems, our finding highlights the importance of using a local probe together with detailed modeling to check thoroughly for crystal imperfections that may give rise to a trivial ZBCP unrelated to Majorana physics.
\end{abstract}

\maketitle

The search for Majorana fermions has become a research priority in recent years because of potential applications in topologically protected quantum computing \cite{Bravyi2000,Bravyi2006,Alicea2011}. It has been proposed that Majorana fermions can be created at the interface between an $s$-wave superconductor and a topological material \cite{Fu2008}. The signature of the Majorana fermion \cite{Law2009,Akhmerov2011} would be a zero bias anomaly in the tunneling conductance at the interface. Therefore, zero bias spectral anomalies observed in a number of experiments \cite{Deng2012,Mourik2012,Das2012,Xu2015} have drawn much attention. However, there are many other possible origins of the zero bias anomalies, such as disorder \cite{Liu2012} and Kondo effects \cite{Cho2015}. Therefore, extreme care is required to interpret and verify the origins of any observed zero bias anomaly \cite{JiaPRL2014, SunPRL2016, YinNatPhys2014, WangScience2018}.

Topological semimetal Sb(111) is a promising platform for interfacing with a superconductor, due to its clean, stable surface and long-lived surface states \cite{Soumyanarayanan2015}. Because of its simple bilayer structure, it has a smaller critical thickness than other topological materials to decouple top and bottom surface states \cite{Yao2013}. Moreover, it is less fragile \cite{Park2015} than tetradymite topological insulators, so a superconducting film could be grown on top without destroying it. In addition, its semiconducting bulk screens chemical potential variations \cite{Roushan2009,Beidenkopf2011}, leading to a more homogeneous surface state, more suitable for superconductor proximity effect study.

Here we show that a robust zero bias conductance peak (ZBCP) can appear on the clean cleaved surface of Sb(111), without proximity to superconductivity. Through a combination of quasiparticle interference imaging, Landau level spectroscopy, and density functional theory, we demonstrate that the ZBCP originates from a sub-surface stacking fault: an isolated single layer of Sb, buried within the normal bilayer structure. The broken layer introduces additional states at negative energy, pushing a saddle point up towards the Fermi energy and producing a van Hove singularity at the Fermi level. We thus introduce a cautionary note in the search for Majorana fermions at Sb-superconductor interfaces, and more generally we exemplify how a trivial crystal defect can robustly mimic the long-sought Majorana.

High-purity antimony (99.999$\%$, from Alfa Aesar\textsuperscript{\textregistered}) in shot form (10.15 g, 6 mm) was sealed in an evacuated quartz tube, and heated in a box furnace to $700^\circ$C for 24 hours. The furnace was cooled slowly ($0.1^\circ$C/min) to $500^\circ$C, and subsequently cooled to room temperature. Scanning tunneling microscope (STM) measurements were performed using a home-built STM at liquid helium temperatures. Single crystals of Sb were cleaved \textit{in situ} in cryogenic ultra-high vacuum to expose the (111) face, and inserted into the STM. Mechanically cut PtIr tips, cleaned by field emission and characterized on gold, were used for the measurements. Spectroscopy data were acquired using a lock-in technique at 1.115 kHz, and conductance maps were obtained by recording out-of-feedback $dI/dV$ spectra at each spatial location.

\begin{figure}[t]
\includegraphics[scale=1]{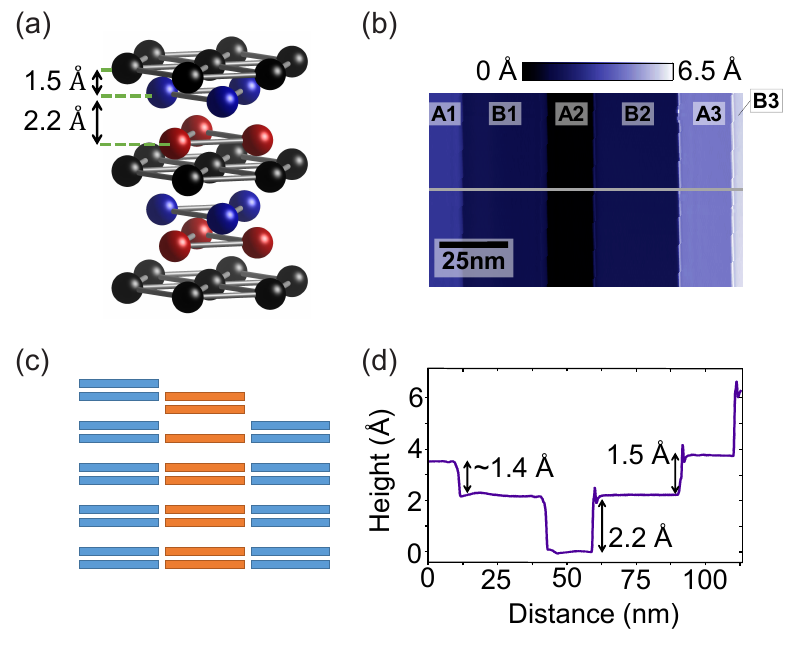}
\caption{\label{fig1}\textbf{Terrace Structure of Antimony.} \textbf{(a)} Bilayer crystal structure of Sb(111). Intrabilayer distance is 1.5 \AA\ and interbilayer distance is 2.2 \AA. \textbf{(b)} Topography of steps observed on Sb(111) by scanning tunneling microscope (STM). (Sample bias, $V_0=300$ mV; junction resistance, $R_J$=3 G$\Omega$.) 
\textbf{(c)} A cartoon model showing how the observed step heights can arise from bilayers and a single broken layer of Sb. The blue terrace is called `normal' and the orange terrace with a single broken layer is called `abnormal'. \textbf{(d)} Height profile perpendicular to the steps, along the grey line in (b). From these step heights, terraces B1, B2 and B3 in (b) can be identified as abnormal.}
\end{figure}

The structure of Sb(111) is shown in Fig.~\ref{fig1}(a), from which we can see that the inter-bilayer distance is 2.2 \AA\ and the intra-bilayer distance is 1.5 \AA\ \cite{Barrett1963,Liu1995}. Since the bonding is much stronger within a bilayer than between bilayers, we expect that the sample will typically terminate between bilayers. The expected step height on terraced surfaces, 2.2 \AA\ + 1.5 \AA\ = 3.7 \AA, is indeed observed in most studies \cite{Gomes2009,Yao2013,SoumyanarayananArxiv2013}. However, Fig.~\ref{fig1}(b) shows a topographic image of an unusual terraced region of the cleaved surface of Sb(111). The height profile perpendicular to the steps is shown in Fig.~\ref{fig1}(d).  There are two types of steps, with heights around 1.5 \AA\ and 2.2 \AA. These are much closer to the intra- and inter- bilayer distances, respectively, than to the expected full-unit-cell step height of 3.7 \AA. These unusual step heights can arise from the configuration shown in Fig.~\ref{fig1}(c), where there is a single broken layer at or beneath the surface of a terrace. The single broken layer could be due to the cleavage process or a pre-existing stacking fault within the sample. In the following, we will refer to this kind of unusual terrace as an `abnormal' terrace and the terrace without a single broken layer as a `normal' terrace. From the height profile in Fig.~\ref{fig1}(d), terraces B1, B2, and B3 in Fig.~\ref{fig1}(b) can be identified as abnormal.

\begin{figure}[t]
\includegraphics[scale=1]{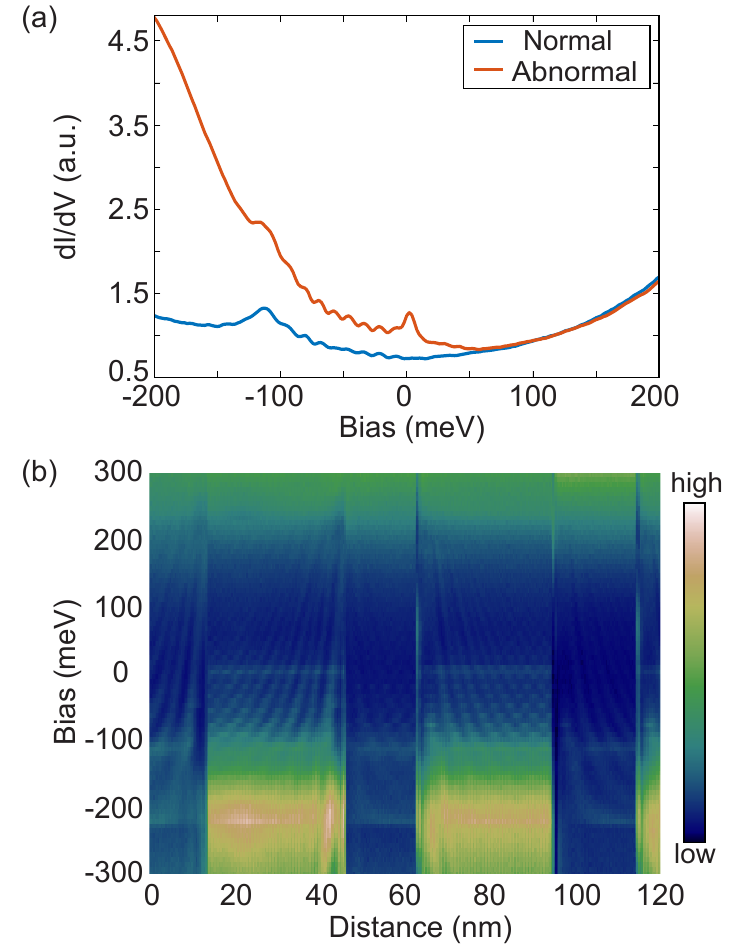}
\caption{\label{fig2}\textbf{$\bm{dI/dV}$ spectra on normal and abnormal terraces.} \textbf{(a)} Averaged $dI/dV$ spectra on the normal (blue) and abnormal (orange) terraces, featuring the zero bias conductance peak (ZBCP) on the abnormal terrace. They are averaged along the grey line on the A2 normal terrace and B1 abnormal terrace in Fig.~\ref{fig1}(b) respectively. ($V_0$ = 300 mV; $R_J$ = 300 M$\Omega$; $V_{\mathrm{rms}}$ = 5 mV.) \textbf{(b)} A set of $dI/dV$ spectra (arbitrary units) acquired along the grey line in Fig.~\ref{fig1}(b). The color depicts the magnitude along each $dI/dV$ spectrum. In addition to the robust ZBCP on the abnormal terraces, the spectra show quasiparticle interference patterns on all terraces. ($V_0$ = 300 mV; $R_J$ = 300 M$\Omega$; $V_{\mathrm{rms}}$ = 5 mV; $T=5.4$ K.)}
\end{figure}

Fig.~\ref{fig2}(a) displays averaged spectra on the normal and abnormal terraces in Fig.~\ref{fig1}(b). The abnormal terraces show a robust ZBCP, with width $\sim$13 meV from a Lorentzian fit. Individual $dI/dV$ spectra, acquired along the grey line in Fig.~\ref{fig1}(b), are shown in Fig.~\ref{fig2}(b), where color depicts the relative intensity of $dI/dV$, which is proportional to the density of states (DOS). The zero bias peaks are quite robust on all three abnormal terraces shown. There is also a series of peaks in each $dI/dV$ curve, equally spaced in energy, from approximately -100 to 0 meV. These quantized peaks have been previously identified on terraced Sb(111) as arising from partial quantum confinement of surface states on finite terraces \cite{Seo2010} (see supplementary section 1).

First we consider whether there is any possibility of superconductivity near our Sb(111) surface, which could give rise to a Majorana mode via predicted mechanisms \cite{Fu2008}. Sb can superconduct with critical temperature $T_c=3.5$ K at high pressure \cite{Wittig1969}, but the temperature in our experiments ranged from 4.5 to 5.4 K. A metastable highly-strained state of Sb with a collapsed face-centered cubic structure may have $T_c$ as high as 7.5 K \cite{Reale1978}. However, we observe no significant change of the lattice constant on the abnormal surface (see supplementary section 2), which implies that there is no such crystal structure change leading to superconductivity in our sample.

\begin{figure}[t]
\includegraphics[scale=1]{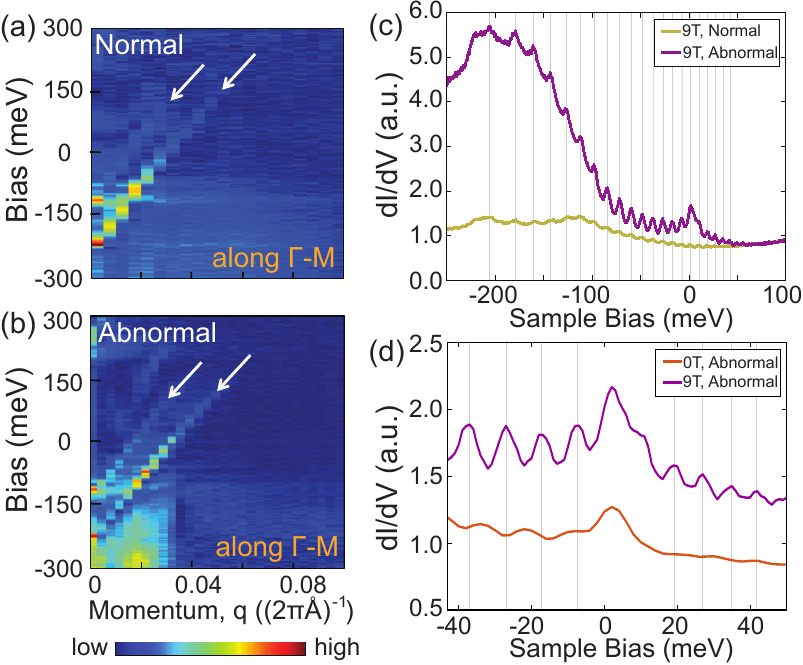}
\caption{\label{fig3}\textbf{Quasiparticle Interference and Landau Quantization.} \textbf{(a,b)} Representative 1D Fourier Transforms of the $dI/dV$ linecuts from the normal (a) and abnormal (b) terraces from Fig.~\ref{fig1}(b), showing two similar prominent dispersing modes along the $\Gamma$-M direction. \textbf{(c)} Comparison of $dI/dV$ spectra at magnetic field $B=9$ T on the abnormal terrace (purple line, $V_0=300$ mV; $R_J=300$ M$\Omega$; $V_{\mathrm{rms}}=1$ mV; $T=4.5$ K) and normal terrace (brown line, $V_0=100$ mV; $R_J=0.1$ G$\Omega$; $V_{\mathrm{rms}}=0.4$ mV; $T=2.2$ K). The vertical grey lines demonstrate that the Landau levels peak at the same energies on the normal and abnormal terraces. The curves are normalized in the positive energy end. \textbf{(d)} Zoomed-in comparison of the ZBCP on the abnormal terrace at $B=9$ T (purple line, $T=4.5$ K) and $B=0$ T (orange line, $T=5.4$ K). The purple curve is offset for clairfy. The ZBCP appears unchanged by the magnetic field. The grey lines indicating LL peaks of the purple line demonstrate that the wigglings on the orange line are different from the LL peaks.}
\end{figure}


To clarify the origin of the ZBCP on the abnormal terrace, we investigate the momentum-resolved electronic structure on this terrace. There are two ways STM can obtain this information, through quasiparticle interference (QPI) and Landau level (LL) spectroscopy, which can be obtained simultaneously on Sb \cite{Soumyanarayanan2015}. QPI is the process by which states of initial wavevector $\vec{k}_i$ and energy $\varepsilon$ scatter elastically and interfere with outgoing states of wavevector $\vec{k}_f$, producing a standing wave pattern with wavevector $\vec{q} = \vec{k}_{f} - \vec{k}_i$ in the $dI/dV$ map at energy $\varepsilon$. The measured $\vec{q}(\varepsilon)$ can be inverted to find the band structure $\vec{k}(\varepsilon)$. LLs occur when an applied magnetic field $B$ quantizes the electronic density of states (DOS), resulting in oscillations in $dI/dV$, whose peak energies can be fit to determine $\vec{k}(\varepsilon)$.

The QPI standing waves on the terraces in Fig.~\ref{fig2}(b) arise from scattering of surface states by the step edges. Figure~\ref{fig3}(a-b) show the 1D Fourier transform of the QPI patterns on the normal and abnormal terraces, respectively. Both terraces show two prominent linear dispersions, starting from around -200 meV and -100 meV, corresponding to two scattering modes in the double-cone surface state band structure of Sb, in agreement with previous studies \cite{Seo2010,Yao2013,Soumyanarayanan2015}. The slopes of the two dispersions on the normal and abnormal surfaces are similar from linear fit, as shown in Table \ref{tab:QPIdispersions}. Apparently the main features of the surface states of the abnormal terrace are quite similar to those of the normal terrace.

\begin{table}[h]
\begin{tabular}{|l|rcl|rcl|}
& \multicolumn{3}{|c|}{Normal} & \multicolumn{3}{|c|}{Abnormal} \\
\hline
Upper & 7.0 & $\pm$ & 0.4 & 7.5 & $\pm$ & 0.2 \\
Lower & 6.8 & $\pm$ & 0.2 & 7.1 & $\pm$ & 0.1 \\
\end{tabular}\protect\caption{Measured slopes from QPI dispersion, in eV$\cdot$\AA.\label{tab:QPIdispersions}}
\end{table}

The other phenomena that can reflect the band structure is LLs in the $dI/dV$ spectrum. As in other topological materials, we can interpret the LLs in the Dirac fermion picture \cite{Hanaguri2010,Okada2011,Cheng2010,Jiang2012}, wherein the energy of the $n^{\mathrm{th}}$ LL $\varepsilon_n$ increases with $\sqrt{nB}$. The Bohr-Sommerfeld quantization relation suggests the momentum space radius $k_n$ for the $n^{\mathrm{th}}$ LL orbit is also proportional to $\sqrt{nB}$ \cite{Hanaguri2010}. So, by tracing the energy of the $n^{\mathrm{th}}$ peak in the DOS spectrum, we can deduce the dispersion of the energy of surface states versus momentum. Fig.~\ref{fig3}(c) demonstrates that the LL peaks on normal and abnormal terraces of Sb, at the same $B=9$ T, match quite well. This also suggests that the band structure of the surface states on the abnormal terrace are quite similar to that on the normal terrace. Fig.~\ref{fig3}(d) shows that at $B=9$ T, the ZBCP is still robust, and its width remains around 10 meV by the Lorentzian fit. The absence of peak splitting means the possibility for Kondo effect \cite{Niu2015,Cho2015} to play a role in inducing the ZBCP is not apparent from the experiment. Otherwise, an energy splitting of $\Delta E \approx$13.4 meV would be expected from Kondo resonance with $\Delta E=2g\mu_BB$ \cite{Liu2013}, given g$\approx$12.8 for Sb \cite{SoumyanarayananArxiv2013}.

\begin{figure}[t]
\includegraphics[scale=1]{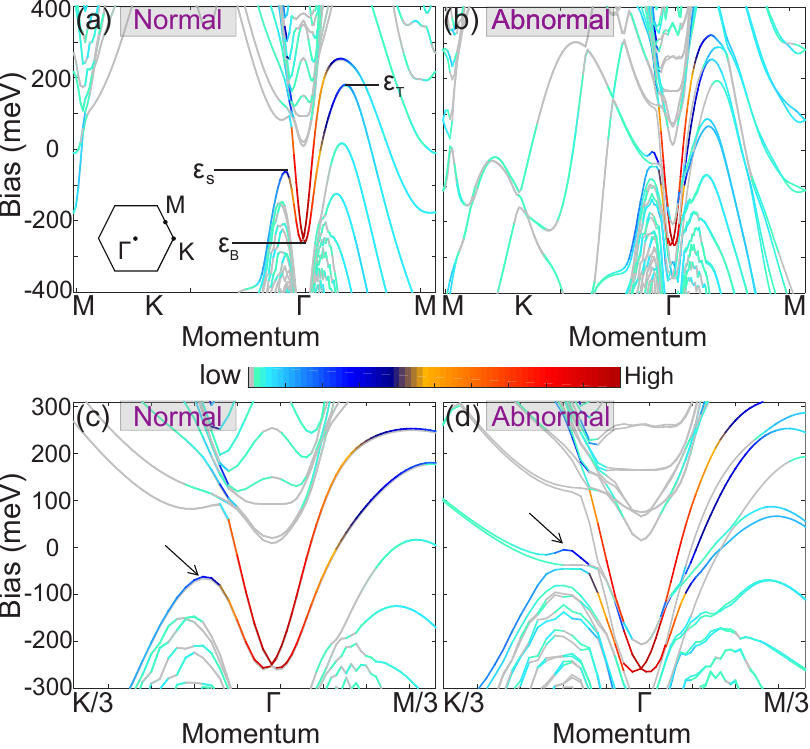}
\caption{\label{fig4}\textbf{Band Structure of normal and abnormal terraces.} \textbf{(a,b)} Band Structure of the normal (a) and abnormal (b) terraces from density functional theory (DFT). The color intensity is the projection of the $p_z$ orbital content at the top surfaces. States that are unavailable to the STM tip appear in grey. \textbf{(c,d)} Zoomed band structures of the normal (c) and abnormal (d) surfaces around the $\Gamma$ point. The saddle points are indicated by the arrows.}
\end{figure}

To better understand the origin of the ZBCP, we performed density functional theory (DFT) calculations using VASP, with Perdew-Burke-Ernzerhof (PBE) pseudopotential. We numerically simulated the band structures and surface states for the normal and abnormal terraces, using a slab geometry. The results along the M-K-$\Gamma$-M directions are plotted in Fig.~\ref{fig4}(a) to (b). STM tunneling current measures the overlap between tip and surface wavefunctions. The outermost valence electrons in Sb are in $s$ and $p$ orbitals. Since the Sb $p_z$ orbital points out of the surface, and reaches farther towards the tip than the Sb $s$ orbital, the $p_z$ orbital is expected to dominate the STM measurement. Hence, to recognize the main contributor in experiment, we projected the available electronic states in $p_z$ orbital on the surface in the band structure. The grey lines in the figures are the unavailable states on the surface. The available states are colored according to their normalized relative contribution to the DOS spectrum. In addition, a scaling factor of 1.2 was applied in energy in order to better match the experimental features on the normal terrace. The same energy scaling factor was employed on the abnormal terrace. But the energy scaling would not affect the energy value of the feature at 0 meV giving rise to the zero bias peak.

For the normal terrace in Fig.~\ref{fig4}(a), the surface state is a double Dirac cone ranging from around -250 meV to 100 meV. There is a saddle point ($\varepsilon_S$) at around $-100$ meV. The energy of the minimum of the cones ($\varepsilon_B$) are a bit above -300 meV and that of an extreme $\varepsilon_T$ is just above 200 meV. The shapes and main features of the calculated band structure for normal terrace thus agree quite well with previous experiment \cite{Soumyanarayanan2015}. The subtle difference in the exact value could be due to the finite size of the slab (46 layers, or 23 bilayers in total were used for the normal terrace).

For the calculation on the abnormal terrace, we started with the single broken layer at the surface of the slab. However, after DFT relaxation of the structure, the single broken layer sunk into the bulk of the slab. The calculation indicates that even if the single broken layer was created at the surface during the \textit{in situ} cleaving process, other layers would rearrange themselves so that the single broken layer would appear below the surface in order to lower the energy. The structure can be stablized when the broken layer is at more than 4 bilayers below the surface. As shown in supplementary section 3, the energy is generally lowered when the single broken layer is at 5 to 8 bilayers below the surface. Also, when the broken layer is 4 bilayer below the surface, the states from the broken layer interfere too much with the surface such that no clear feature of surface state can be recognized. So we did our calculation mostly at cases where the broken layer is a bit farther away from the surface, but not too far such that the effect of the broken layer is still apparent. Fig.~\ref{fig4}(b) shows the typical result for the abnormal terrace, using a slab of 47 layers with the single broken layer 6 bilayers below the surface.

The contribution of $\vec{k}$ states to the STM-measured $dI/dV$ is expected to decay exponentially away from the $\Gamma$ point \cite{Tersoff1983}. So we focus on the band structure around the $\Gamma$ point as shown in Fig.~\ref{fig4}(c) and (d) for the normal and abnormal surfaces. Apparently, the double-cone feature of the normal surface state is preserved in a similar energy range on the abnormal terrace. This explains why similar QPI and LL quantization, which reflect the characteristics of the surface state, were observed on the abnormal terrace in our experiment.

Now we turn our attention to the saddle point, and show that it is responsible for the ZBCP on the abnormal surface. As mentioned above, on the normal terrace, there is a saddle point ($\varepsilon_S$) around -100 meV on the outer cone along the $\Gamma$-K direction, which causes a peak at around -100 meV in the DOS spectrum of the normal terrace \cite{Soumyanarayanan2015}. However, on the abnormal terrace (Fig.~\ref{fig4}(e)), there are more states appearing below this saddle point, pushing it up to the Fermi energy (0 meV). This will constitute a van Hove singularity leading to a peak in the zero bias of the DOS spectrum. This pushed-up saddle point can be seen in slabs with the single broken layer at different depths (as long as it can be stabilized). Other plots of band structure for the abnormal terrace with slab of different thickness and single broken layer at different depth can be also be found in the supplementary section 4. So our findings suggested that the origin of the zero bias peak observed in experiment is a trivial (non-topological) effect due to the sub-surface stacking fault: the single sunken broken layer.

Other characteristics in the DOS spectrum of the abnormal terrace from Fig.~\ref{fig2}(a) included a larger bump at negative energies from -200 meV to 0 meV,  compared with that of the normal terrace. Comparing the band structures of the normal (Fig.~\ref{fig4}(c)) and abnormal (Fig.~\ref{fig4}(d)) terraces, there are more available states on the abnormal terrace around the $\Gamma$ point from around -200 meV to 0 meV. This naturally explains why the $dI/dV$ conductance is enhanced on the abnormal terrace from -200 meV to 0 meV.

In conclusion, we reported that the ZBCP, which is the signature of Majorana fermion, can arise trivially from a sub-surface stacking fault within topological material, even if the surface is clean. Amidst the sprint to stake claims on new Majorana fermion systems, our finding highlights the importance of using a local probe and detailed modeling to check thoroughly for crystal imperfections that may give rise to a trivial ZBCP unrelated to Majorana physics.

\bibliography{Sb_ZBP}

\section{acknowledgements}
\begin{acknowledgements}
This work is supported by the National Science Foundation under Grant No. DMR-1410480, and the Canada Excellence Research Chair program. J.E.H. acknowledges support from the Canadian Institute for Advanced Research.
\end{acknowledgements}


\clearpage
\widetext
\begin{center}
	\textbf{\large Supplementary Materials}
\end{center}

\setcounter{equation}{0}
\setcounter{figure}{0}
\setcounter{table}{0}
\setcounter{page}{1}
\makeatletter
\renewcommand{\theequation}{S\arabic{equation}}
\renewcommand{\thefigure}{S\arabic{figure}}
\renewcommand{\bibnumfmt}[1]{[S#1]}

\section{Section 1}
\label{sec1}
\begin{center}
\textbf{Spatially Quantized Resonances in Zero-Field $\bm{dI/dV}$ spectra}
\end{center}
\begin{figure}[h]
\includegraphics[scale=1.2]{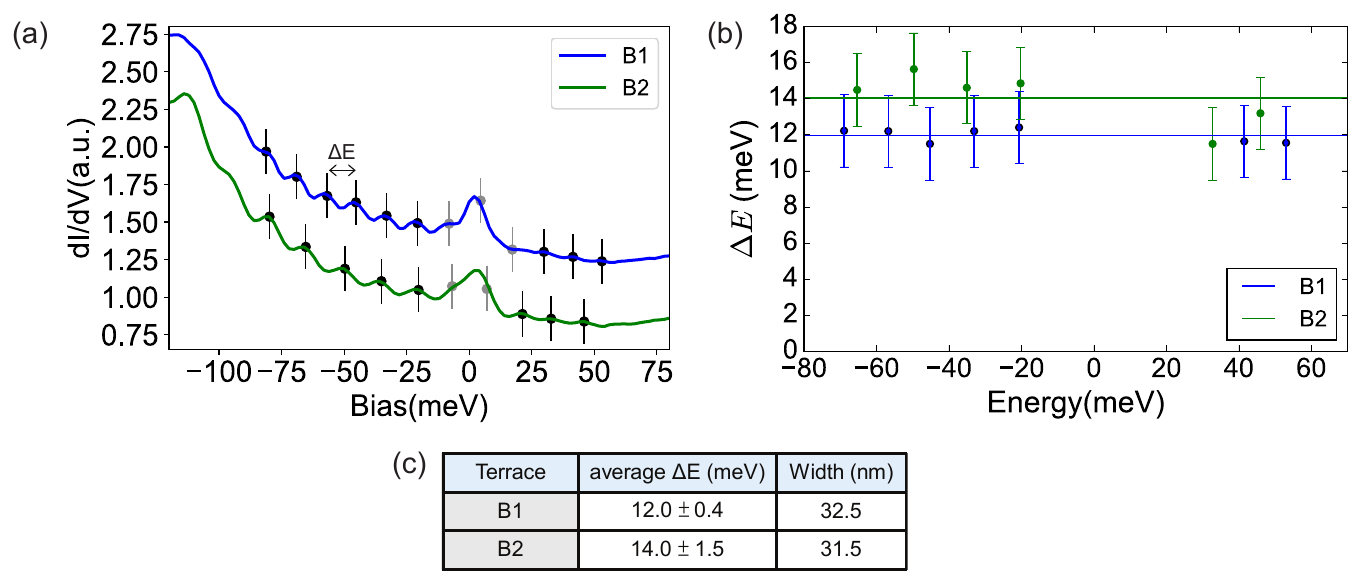}
\caption{\label{supp1}\textbf{$\bm{dI/dV}$ spectra on abnormal surfaces and the energy difference between peaks} \textbf{(a)} Averaged $dI/dV$ measurement on the B1 and B2 terraces in Fig.~\ref{fig1}(b), where the most obvious quantized peaks can be observed. $\Delta E$ is defined to be the energy difference between adjacent peaks. The spectra are offset vertically for clarity. (Data acquired with $V_0=300$ mV; $R_J=300$ M$\Omega$; $V_{\mathrm{rms}}=5$ mV, 2 meV energy spacing.) The black vertical lines are indicating the wiggling peaks that can be fitted by a Gaussian after polynomial subtraction of the curve and whose energy positions are used for calculating $\Delta E$ to be shown in (b). The grey vertical lines are equally spaced around the ZBCP showing the expected position of wiggling peaks in this region. \textbf{(b)} Energy spacing between quantized peaks in (a), quantified by Gaussian fitting after polynomial subtraction of the curves. The horizontal lines show the average energy separation for each terrace. \textbf{(c)} A table showing the widths and average $\Delta E$ of terrace B1 and B2 with the standard derivation.}
\end{figure}

As demonstrated in Fig.~\ref{supp1}(b), the energy difference between the peaks in the $dI/dV$ spectra in Fig.~\ref{supp1}(a) is quite constant on the B1 and B2 terraces, given that the sampling rate in Fig.~\ref{supp1}(a) is 2 meV per energy point and the lock-in $V_{\mathrm{rms}}$ amplitude is 5 meV. The average $\Delta E$ is roughly inversely proportional to the width of the step as expected from ~\cite{Seo2010} where similar peaks were observed and explained by quantized resonances arising from boundary conditions on finite terraces of Sb(111).

On the abnormal steps B1 and B2, the amplitude of the ZBCP is much larger than the the peaks from spatially quantized resonances. Similar enhanced peak at zero energy was not seen in ~\cite{Seo2010}. Also, we attempted to plot the expected positions of wiggling peaks in the part of the curve around the ZBCP by the grey vertical line in Fig.~\ref{supp1}(a). It can been seen that the ZBCP is deviated from the expected position of the wiggling peak, especially on the terrace B2. These are suggesting that the ZBCP is not from the quantum resonance of the steps.

\newpage
\section{Section 2}
\label{sec2}
\begin{center}
\textbf{Justification of Having No Crystal Structure Change on the Abnormal Surface}
\end{center}
\begin{figure}[h]
\includegraphics[scale=1]{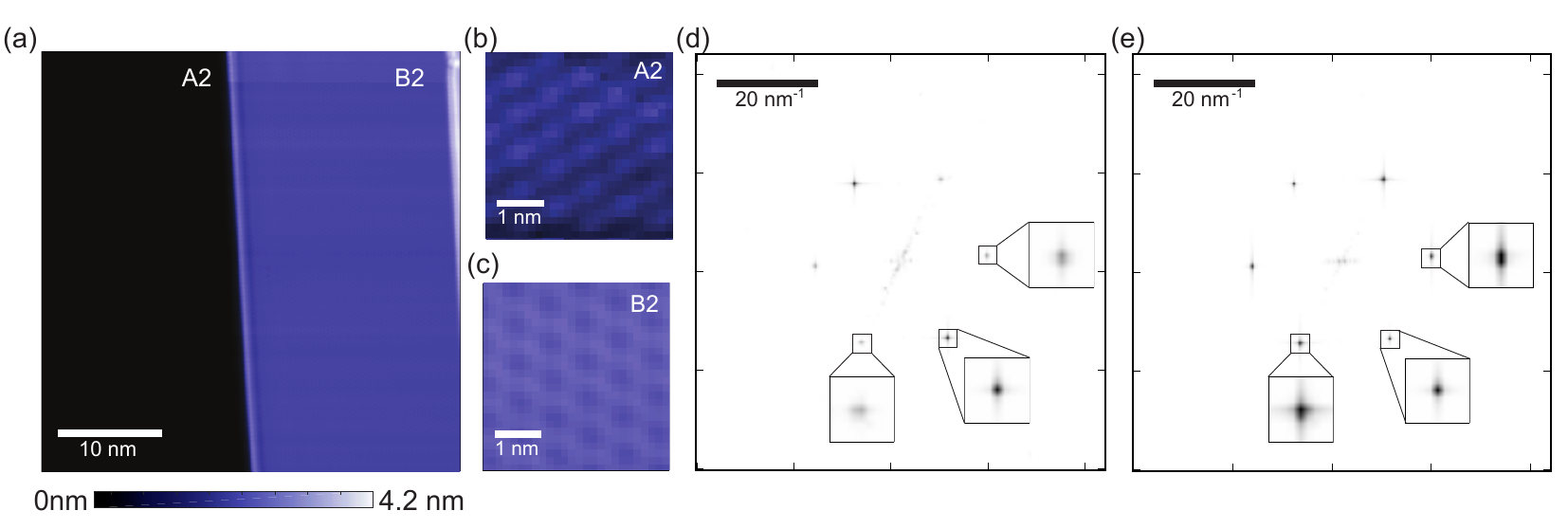}
\caption{\label{supp2}\textbf{STM Topography of the normal and abnormal steps and its Fourier transform} \textbf{(a)} STM Topography of the step with the normal surface A2 on the left side and abnormal surface B2 on the right side. The total field of view is 40$\times$40 nm. ($V_0=300$ mV; $R_J=3$ G$\Omega$; resolution = 512$\times$512 pixels) \textbf{(b,c)} Cropped images from surface (b) A2 and (c) B2 showing atomic resolution. \textbf{(d) \& (e)} Fourier transforms of (d) normal and (e) abnormal surfaces of equal square size (210 pixels) cropped from (a). We can see that the hexagonal Bragg peaks are at the same momentum location on both the normal and abnormal surface. This reveals that there is no lattice constant change on the abnormal surface.}
\end{figure}

Sb in its usual A7 crystal structure has been shown to superconduct under high pressure of 85 kbar with a critical temperature of $T_c=3.5$ K \cite{Wittig1969}. All data shown in this work was acquired at higher temperatures between 4.5 and 5.4 K (except for one spectrum acquired at $B=9$ T and $T=2.2$ K in Fig.~\ref{fig3}(d)), so the high-pressure A7 superconducting state is not relevant to our work. But superconductivity has also been demonstrated up to $T_c=7.5$ K in a metastable A1 state that can be stabilized by epitaxial strain in Sb films on MnO \cite{Reale1978}.

The distance between atoms on the usual Sb(111) surface of the A7 structure is 4.301 $\AA$ at $T=4.2$ K \cite{Liu1995}. In the metastable state of Sb of FCC structure with critical temperature above 7K suggested by Reale \cite{Reale1978}, the distance between atoms in hexagonal pattern on the (111) surface would become about 3.32 \AA, around 25$\%$ smaller than the usual distance. This would give us a set of Bragg peak in the fourier space with 1/3 larger magnitude corresponding to that shortened atomic distance.

However, as we demonstrated in Fig.~\ref{supp2}, when we Fourier transform the normal and abnormal surfaces respectively, if the abnormal surface had a different lattice constant as the normal surface, we would expect hexagonal Bragg peaks in the Fourier space of the abnormal surface are at momentum about 1/3 larger than the momentum of the Bragg peaks on the normal surface. However, as apparent in the comparison between Fig.~\ref{supp2}(d) and (e), the hexagonal Bragg peaks on the normal and abnormal surfaces are at similar momentum. This show that there is no crystal structure change on the abnormal surface and thus there would not be potential for the abnormal surface to be superconducting at our operating temperature. In this way, we can verify that there is no superconductivity involved in the ZBCP on the abnormal terrace.

\newpage
\section{Section 3}
\label{sec3}
\textbf{Relaxed Energy of Sb slabs of 47 total layers with a single broken layer at different bilayer below the surface (different depths)}
\begin{figure}[h]
	\includegraphics[scale=0.75]{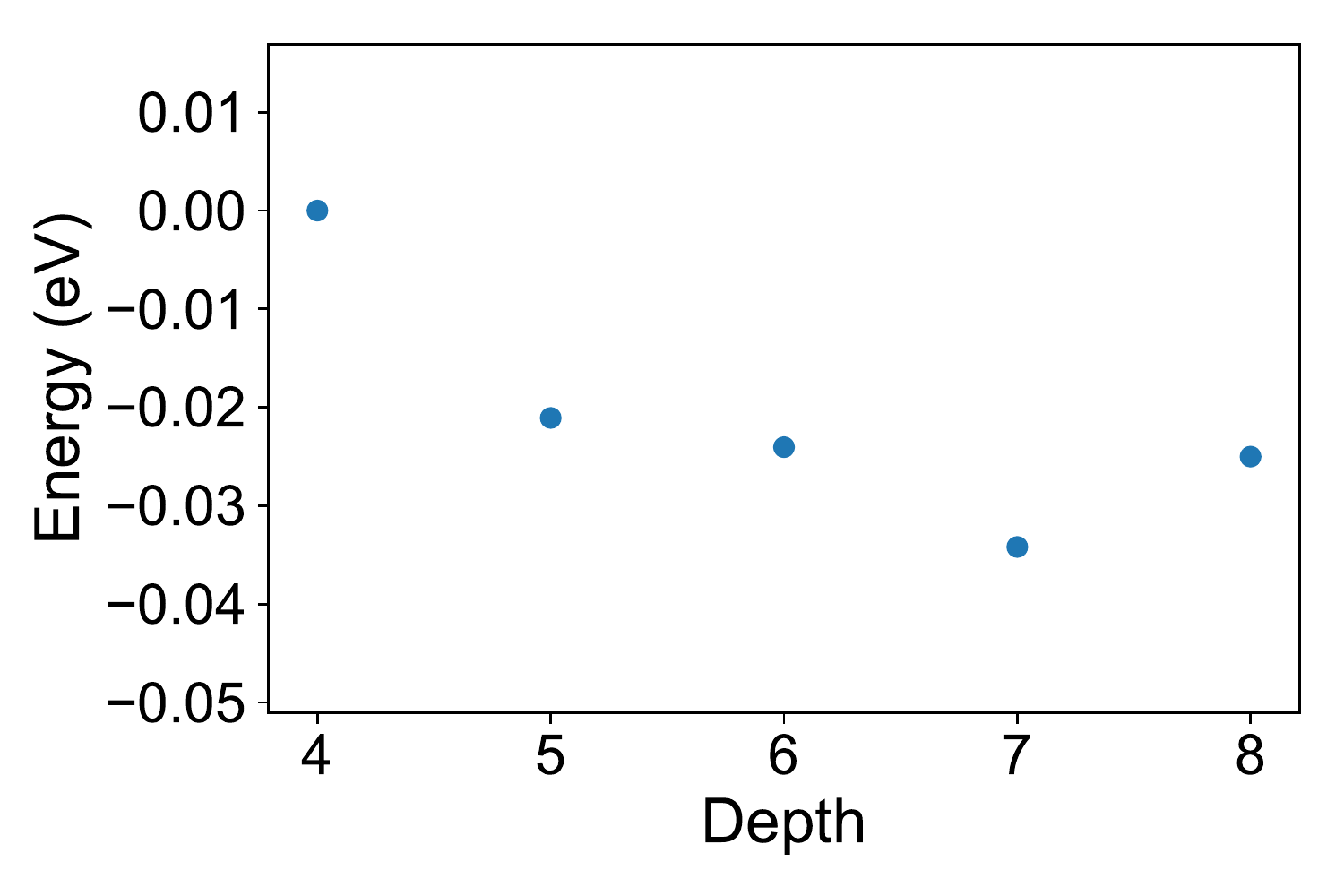}
	\caption{\label{supp3}\textbf{Total energy of slabs with the single broken layer at different bilayer below the surface after DFT relaxation} In the horizontal axis, 'Depth = $i$' means the single broken layer is $i$ bilayers below the surface. The energy of Depth 4 is set to be the reference energy. It can be seen that Depth 5 to 8 generally have lower energy than Depth 4.}
\end{figure}

\newpage
\section{Section 4}
\textbf{Band Structure from DFT with different slab thickness and depth of the single broken layer}
\begin{figure}[h]
\includegraphics[scale=1]{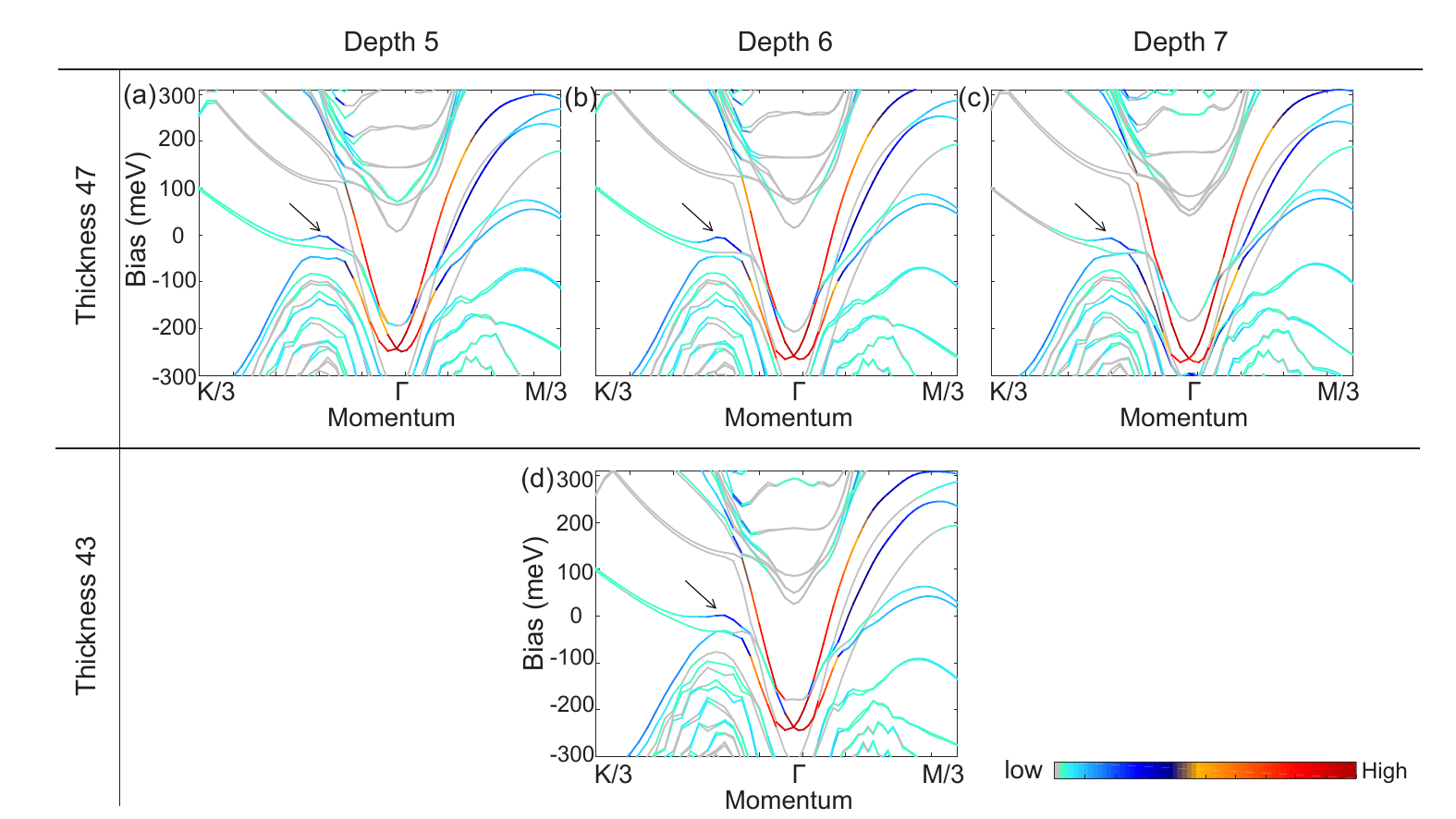}
\caption{\label{supp4}\textbf{Band structures from DFT with different thickness and depth of the single broken layer} \textbf{(a-c)} Band structures from DFT using slabs with 47 total layers and single broken layer at depth 5 (a), 6 (b) and 7 (c). They all have a saddle point (indicated by the black arrow) around 0 meV. \textbf{(d)} Band structure from a slab with different thickness (43 layers in total) and single broken layer at depth 6. Ideally the larger the slab, the smaller the finite size effect and the more reliable is the calculation. (And 47 total layers is the limit of our computational resources.) But comparing (b) and (d), we can still identify similar main features of the band structures of the abnormal surfaces, especially the saddle point around 0 meV.}
\end{figure}

\end{document}